\documentclass[twocolumn,english,aps,floatfix,superscriptaddress,amsfonts,amssymb,superscriptaddress]{revtex4}
\usepackage{color}
\usepackage{graphicx}
\usepackage{amsmath}
\usepackage{latexsym}
\usepackage{epstopdf}
\usepackage{amsmath,amssymb}
\usepackage{amssymb,amsmath}
\usepackage{multirow}
\usepackage{mathtools}
\usepackage{bbold}
\usepackage[dvipsnames]{xcolor}
\newcommand{\beq}{\begin{equation}}
\newcommand{\eeq}{\end{equation}}

\usepackage{tikz}
\usepackage{xcolor}
\usetikzlibrary{patterns}

\begin{document}

\title{Full counting statistics  of the subsystem energy for free fermions and quantum spin chains}

\author{K. Najafi}
\affiliation{ Department of Physics, Georgetown University, 37th and O Sts.  NW, Washington, DC 20057, USA}

\author{M.~A.~Rajabpour}
\affiliation{  Instituto de F\'isica, Universidade Federal Fluminense, Av. Gal. Milton Tavares de Souza s/n, Gragoat\'a, 24210-346, Niter\'oi, RJ, Brazil}

\date{\today{}}

\begin{abstract}
We calculate the full counting statistics (FCS) of a subsystem energy in  free fermionic systems by means of the Grassmann variables. We demonstrate that the generating function of these systems can be written 
as a determinant formula with respect to the Hamiltonian couplings and by employing the Bell's polynomials, we derive exact formulas for the subsystem energy moments. In addition, we discuss the same quantities 
in the quantum XY spin chain, and we demonstrate that at the critical regimes the fluctuations of the energy moments decay like a power-law as we expect from the conformal field theory arguments, while in noncritical 
regimes, the decay is exponential. Furthermore, we discuss the full counting statistics of subsystem energy in the quantum XX chain.

\end{abstract}
\pacs{}
\maketitle
\tableofcontents

\section{Introduction}
\setcounter{equation}{0}
\renewcommand{\theequation}{I\arabic{equation}}

In quantum many-body physics, the fluctuations of local observables carry a fair amount of information regarding the physical properties of the system. The rapid progress in the manipulation of quantum devices in condensed
matter physics has made it possible to investigate the
nature of fluctuations in quantum many-body systems to even higher degrees of details. In quantum
mechanics, one usually is interested in the first few moments of the fluctuations while the full description of the system requires
the knowledge of the full distribution function of the observable. These kinds of distribution functions are examples of a more general concept known as full counting statistics (FCS) which studies the full distribution of a
macroscopic
observable in arbitrary systems. The term FCS was first popularized in the study of charge
transport in mesoscopic systems\cite{Levitov1993,Blanter2000,
Levitov2005,Sukhorukov2007,Belzig2005,Nazarov2007,Kaasbjerg2015,Dasenbrook2016,Szczepanski2016}. 
However, since this quantity can 
be defined naturally in any quantum system, it has been also applied in different areas such as Fermi edge problems\cite{Abanin2005},
ultracold atoms\cite{Gritsev2006,Polkovnikov2006,Hofferberth,Calabrese2}, quantum chains  
\cite{Klich2009,Klich2012a,Klich2012b,Abanov2013,Chiara2016,Stephan2017,Essler2017}, 
 and many-body localization\cite{Luitz2015,Pollmann2016}. For a good list of applications and references, 
see Refs.\cite{Klich2012a,Klich2014}. In the context of quantum chains, 
one can simply consider
the system in its ground state and then study the full distribution of
the number of fermions \cite{Klich2009,Klich2012a,Klich2012b} or magnetization in the subsystem
\cite{Eisler2003,Cherng2007,LF2008,Abanov2013,Chiara2016,Stephan2017,Essler2017}
and find the relevant universal scaling functions\cite{Goldenfeld2001,Bruce1981,Binder1981}. One of the natural
quantities that can be defined for any system is the energy of the subsystem. The Hamiltonian truncated to a subsystem does not commute with the Hamiltonian of the full system. Consequently,
if the full system is in its ground state, the energy of the subsystem can have many different values with different probabilities.
Surprisingly, to the best of our knowledge, the FCS of the subsystem energy has not been studied in the literature.
However, in Ref.\cite{Arad2016}, related problems regarding the relation between 
probability distributions of the measurement outcomes of the local and global
Hamiltonians have been discussed. For the distribution of the energy of the full system in the Gibbs state and possible experimental methods to detect it, see Ref.\cite{Galperin2016}. There have also been some works regarding the full counting statistics of both energy transport of phonons \cite{Saito2011,Clerk2010,Clerk2011} and
electron energy transfer\cite{Tang,Seoane,Agarwalla}. In this paper, we study the FCS of the subsystem energy in generic free fermion Hamiltonian with real couplings, and in particular, in quantum spin chains that can be mapped to the free fermions.

To have the full distribution of the system,
one usually needs to calculate the generating function of the distribution functions. In the case of free fermions,
the studies of those observables that have quadratic form in the fermionic representation (for example fermion number) always lead to determinant formulas\cite{Levitov1993,Klich2012b,Klich2014}. 
The  Hamiltonian truncated to a subsystem in free fermions naturally
has a quadratic form and one expects a determinant formula. In this paper, we will find a determinant formula in
a form which is different but equivalent to the ones that can be derived from earlier
approaches, see for example Ref.\cite{Klich2014}. We will implement the fermionic coherent states written in Grassmann representation which is also practical when the state of the system is generic
but has a simple form in the local basis. Afterwards, using the Bell's polynomials, we will provide exact formulas for the moments.
In Sec. III, we will apply our determinant formulas to study the FCS of subsystem energy
in the quantum XY chain and will remark on different properties of this quantity on different regions of the phase diagram. 
In particular, we will study the universal properties in the critical regions. 
 Finally, in Sec. IV, we will summarize our results and future directions.

\section{Full counting statistics of the subsystem energy in the generic free fermions}
\setcounter{equation}{0}
\renewcommand{\theequation}{\arabic{equation}}

In this section, we study the full counting statistics of the subsystem energy in the ground state of a generic real free fermion Hamiltonian in an arbitrary dimension.
To start, we calculate the first two moments and then we provide an exact formula for the generating function of the subsystem energy.
Afterward, we show how one can calculate all of the moments exactly by a proper expansion of the generating function. Consider a generic free fermion Hamiltonian with
the following truncated Hamiltonian for the subsystem $D$;
\begin{eqnarray}\label{HD1}
H_{D}=\textbf{c}^{\dagger}.\textbf{A}.\textbf{c}+\frac{1}{2}\textbf{c}^{\dagger}.\textbf{B}.\textbf{c}^{\dagger}+\frac{1}{2}\textbf{c}.\textbf{B}^{T}.\textbf{c}-\frac{1}{2}{\rm tr}{\textbf{A}},
\end{eqnarray} 
where $\textbf{B}^{T}$ is the transpose of $\textbf{B}$ and $\boldsymbol{c}$ is a vector with elements $c_{i}$ and $i=1,2,...,|D|$, where $|D|$
is the number of sites in the region $D$. Moreover, $\textbf{A}$ and $\textbf{B}$ are 
symmetric and antisymmetric matrices, respectively, to keep the Hamiltonian Hermitian. Note that here we just consider real 
$\textbf{A}$ and $\textbf{B}$.
We are interested in calculating the full counting statistics of energy defined as the following expectation value
\begin{eqnarray}\label{Mlambda1}
M(\bar{\lambda}_{D})=
\langle e^{\bar{\lambda}_{D}H_{D}}\rangle={\rm tr} [\rho_{D}e^{\bar{\lambda}_{D}H_{D}}],
\end{eqnarray} 
where $\bar{\lambda}_{D}=\frac{\lambda}{\sqrt{\langle H_{D}^{2}\rangle}}$. For later implications, it is convenient to introduce new operators defined as,
\begin{eqnarray}\label{New operators}
a_i=c_i^{\dagger}+c_i,\hspace{1cm}b_i=c_i^{\dagger}-c_i,	
\end{eqnarray}
with  the anticommutator relations $\{a_i,a_j\}=2\delta_{ij}$, $\{b_i,b_j\}=-2\delta_{ij}$, and $\{a_i,b_j\}=0$, and the well known correlations defined as block Green matrices,
\begin{subequations}
\begin{eqnarray}\label{Green matrix ba}
G_{ij}^{(ba)} &=&{\rm tr}[\rho_Db_ia_j],\\
G_{ij}^{(ab)} &=&{\rm tr}[\rho_Da_ib_j],\\
G_{ij}^{(aa)} &=&{\rm tr}[\rho_Da_ia_j],\\
G_{ij}^{(bb)} &=&{\rm tr}[\rho_Db_ib_j], 
\end{eqnarray}
\end{subequations}
where $\rho_D$ is the reduced density matrix of the region $D$ and $G_{ij}^{(ab)}=-G_{ji}^{(ba)}$. Note that the role of the Hamiltonian of the full system is encoded in $\rho_D$.   
By substituting operators $a$ and $b$ in the Hamiltonian (\ref{HD1}),   $ H_{D}$ becomes,
\begin{eqnarray}\label{HD2}
H_{D}=\frac{1}{2}[b_{i}(A_{ij}+B_{ij})a_{j}],
\end{eqnarray} 
where we have used $\boldsymbol{G}^{(aa)}+\boldsymbol{G}^{(aa)T}=-(\boldsymbol{G}^{(bb)}+\boldsymbol{G}^{(bb)T})=2\mathbb{1}$. By using the definition of these correlators, we get 

\begin{eqnarray}\label{Hl3}\
\langle H_{D}\rangle &=&\frac{1}{2}{\rm tr}[(\textbf{A}-\textbf{B})\boldsymbol{G}^{(ba)}]=\frac{1}{4}{\rm tr}[\textbf{D}^{T}\boldsymbol{G}^{(ba)}], 
\end{eqnarray} 
where we have introduced a matrix $\textbf{D}=2(\textbf{A}+\textbf{B})$. Similarly, by using the Wick's theorem one can also calculate the $\langle H_{D}^{2}\rangle$. The final result can be written as:

\begin{eqnarray}\label{Hl6}\
\langle H_{D}^{2}\rangle &=&\langle H_{D}\rangle ^{2}+\frac{1}{16}{\rm tr}[\textbf{D}\textbf{D}^{T}]\nonumber \\
&-&\frac{1}{16}{\rm tr}[\textbf{D}^{T}\boldsymbol{G}^{(ba)}\textbf{D}^{T}\boldsymbol{G}^{(ba)}]].
\end{eqnarray} 
It is needless to say that the calculation of the higher moments starts to get cumbersome immediately which makes the forthcoming calculations much more valuable.
In any case, the above two direct calculations can be used to check the validity of the general results.

\subsection{Generating function of the subsystem energy in the ground state}
In this subsection, we come back to our  problem of calculating the generating function of the subsystem energy for a system which is
in its ground state. To do that we can use the fermionic coherent state defined as
\begin{eqnarray}\label{fermionic coherent states1}
 |\boldsymbol{\xi}>= |\xi_1,\xi_2,...,\xi_N>= e^{-\sum_{i=1}^N\xi_ic_i^{\dagger}}|0>,
\end{eqnarray}
where $\xi_i$'s are Grassmann numbers which satisfy the following properties: $\xi_n\xi_m+\xi_m\xi_n=0$ and $\xi_n^2=\xi_m^2=0$. Consequently, we can show
\begin{eqnarray}\label{fermionic coherent states2}
c_i |\boldsymbol{\xi}>= \xi_i  |\boldsymbol{\xi}>.
\end{eqnarray}
By using the Grassmann variables the reduced density matrix can be written  as\cite{Peschela,Barthel2008}
\begin{eqnarray}\label{reduced density matrix xi}
\rho_D(\boldsymbol{\xi},\boldsymbol{\xi'})&=&<\boldsymbol{\xi}|\rho_D|\boldsymbol{\xi'}>\nonumber\\
&=&\det[\frac{1}{2}(\mathbb{1}-\boldsymbol{G}^{(ba)})]e^{\frac{1}{2}(\bar{\boldsymbol{\xi}}-\boldsymbol{\xi}')^T
\boldsymbol{F}(\bar{\boldsymbol{\xi}}+\boldsymbol{\xi}')},
\end{eqnarray}
where we have introduced the matrix $\boldsymbol{F}=(\boldsymbol{G}^{(ba)}+\mathbb{1})(\mathbb{1}-\boldsymbol{G}^{(ba)})^{-1}$. The   trace
in the context of Grassmann variables can be calculated as:
\begin{equation}\label{trace}
{\rm tr}\,\, O=\int d\bar{\boldsymbol{\xi}}d{\boldsymbol{\xi}}e^{- {\bar{\boldsymbol{\xi}}}{\boldsymbol{\xi}}}\langle{-\boldsymbol{\xi}}|O|\boldsymbol{\xi}\rangle.
\end{equation}
Then the  equation (\ref{Mlambda1}) can be written as:
\begin{equation}\label{Mlambda2}\
M(\bar{\lambda}_{D})=
{\rm tr}[\rho_{D}e^{\bar{\lambda}_{D}{H_{D}}}]=\int d\bar{\boldsymbol{\eta}}d{\boldsymbol{\eta}}e^{- {\bar{\boldsymbol{\eta}}}{\boldsymbol{\eta}}}\langle{-\boldsymbol{\eta}}|\rho_{D}e^{\bar{\lambda}_{D}H_{D}}|
\boldsymbol{\eta}\rangle.
\end{equation} 
After using the identity
\begin{equation}\label{identity}\
I=\int d\bar{\boldsymbol{\xi}}d\boldsymbol{\xi}e^{-\bar{\boldsymbol{\xi}}\boldsymbol{\xi}}|\boldsymbol{\xi}\rangle\langle\boldsymbol{\xi}|
\end{equation} 
we get
\begin{eqnarray}\label{Mlambda2}\
M(\bar{\lambda}_{D})=&&\int d\bar{\boldsymbol{\eta}}d{\boldsymbol{\eta}}\int d\bar{\boldsymbol{\xi}}d{\boldsymbol{\xi}}e^{- {\bar{\boldsymbol{\eta}}}{\boldsymbol{\eta}}}
e^{- {\bar{\boldsymbol{\xi}}}{\boldsymbol{\xi}}}\nonumber \\&&\langle{-\boldsymbol{\eta}}|\rho_{D}|\boldsymbol{\xi}\rangle\langle\boldsymbol{\xi}|e^{\bar{\lambda}_{D}{H_{D}}}|\boldsymbol{\eta}\rangle.
\end{eqnarray} 
We can now calculate the two expectations separately. 
To calculate the  second expectation, first we decompose $e^{\bar{\lambda}_{D}{H_{D}}}$ 
by means of the Balian-Brezin formula \cite{Balian1969} as:
\begin{eqnarray}\label{H2}\
e^{\bar{\lambda}_{D}{H_{D}}}=e^{\frac{1}{2}\textbf{c}^{\dagger}\textbf{X}\textbf{c}^{\dagger}}e^{\textbf{c}^{\dagger}\textbf{Y}\textbf{c}}e^{-\frac{1}{2}{\rm tr}\textbf{Y}}e^{\frac{1}{2}\textbf{c}\textbf{Z}\textbf{c}},
\end{eqnarray} 
where $\textbf{X}$, $\textbf{Y}$, $\textbf{Z}$ can be calculated from the blocks of matrix
$\textbf{T}$ defined as 
\begin{eqnarray}\label{mat_T}\
\textbf{T}=e^{\bar{\lambda}_{D} \begin{pmatrix}
\textbf{A} & \textbf{B}\\
\textbf{-B} & \textbf{-A}\\
  \end{pmatrix}  }
 = \begin{pmatrix}
\textbf{T}_{11} & \textbf{T}_{12}\\
\textbf{T}_{21} & \textbf{T}_{22}\\
  \end{pmatrix}.  
\end{eqnarray} 
which leads to
\begin{eqnarray}\label{X}\
\textbf{X}=\textbf{T}_{12}(\textbf{T}_{22}^{-1}),\hspace{0.5cm}\textbf{Z}=(\textbf{T}_{22}^{-1})\textbf{T}_{21},\hspace{0.5cm}e^{\textbf{-Y}}=\textbf{T}_{22}^{T}.
\end{eqnarray} 
By using the above formulas and the properties of the fermionic coherent states, we have
\begin{eqnarray}\label{e_lambda_l}
\langle\boldsymbol{\xi}|e^{\bar{\lambda}_{D}{H_{D}}}|\boldsymbol{\eta}\rangle=
e^{\frac{1}{2}\bar{\boldsymbol{\xi}}\boldsymbol{X}\bar{\boldsymbol{\xi}}+\frac{1}{2}\boldsymbol{\eta}\boldsymbol{Z}\boldsymbol{\eta}-\frac{1}{2}{\rm tr \boldsymbol{Y}}+\bar{\boldsymbol{\xi}}e^{\boldsymbol{Y}}\boldsymbol{\eta}}.
\end{eqnarray}
After implementing the above formula in the equation (\ref{Mlambda2}) we get
\begin{eqnarray}\label{Mlambda3}\
&&M(\bar{\lambda}_{D})=\det[{\frac{1}{2}(\mathbb{1}-\boldsymbol{G}^{(ba)})}]e^{-\frac{1}{2}{\rm tr \boldsymbol{Y}}}\int d\bar{\boldsymbol{\eta}}\int d{\boldsymbol{\eta}}e^{\frac{1}{2}\boldsymbol{\eta}\boldsymbol{Z}
\boldsymbol{\eta}}\nonumber \\ 
&&\int d\bar{\boldsymbol{\xi}}\int d{\boldsymbol{\xi}} e^{-\bar{\boldsymbol{\eta}}\boldsymbol{\eta}-\bar{\boldsymbol{\xi}}\boldsymbol{\xi}+\frac{1}{2}\bar{\boldsymbol{\xi}}\boldsymbol{X}\bar{\boldsymbol{\xi}}
+\bar{\boldsymbol{\xi}}e^{\boldsymbol{Y}}\boldsymbol{\eta}-\frac{1}{2}(\bar{\boldsymbol{\eta}}+\boldsymbol{\xi})^T
\boldsymbol{F}(-\bar{\boldsymbol{\eta}}+\boldsymbol{\xi})}.
\end{eqnarray} 
To calculate the integrals, we first introduce new variables, $\bar{\boldsymbol{\nu}}=\frac{\bar{\boldsymbol{\eta}}+\boldsymbol{\xi}}{\sqrt{2}}$ and 
$\boldsymbol{\nu}=\frac{-\bar{\boldsymbol{\eta}}+\boldsymbol{\xi}}{\sqrt{2}}$, then the integral becomes
\begin{eqnarray}\label{Mlambda4}\
M(\bar{\lambda}_{D})=(-1)^{|D|}\det[{\frac{1}{2}(\mathbb{1}-\boldsymbol{G}^{(ba)})]e^{-\frac{1}{2}{\rm tr \boldsymbol{Y}}}  \int d\boldsymbol{\eta}}\int d{\bar{\boldsymbol{\xi}}}\nonumber \\e^{\frac{1}{2}\bar{\boldsymbol{\xi}}\boldsymbol{X}
\bar{\boldsymbol{\xi}}+\frac{1}{2}\boldsymbol{\eta}\boldsymbol{Z}\boldsymbol{\eta}+\bar{\boldsymbol{\xi}}e^{\boldsymbol{Y}}\boldsymbol{\eta}}\int d\bar{\boldsymbol{\nu}}\int d\boldsymbol{\nu} e^{-\bar{\boldsymbol{\nu}}\boldsymbol{F}\boldsymbol{\nu}-\frac{\bar{\boldsymbol{\xi}}+\boldsymbol{\eta}}{\sqrt{2}}\boldsymbol{\nu}-\frac{\bar{\boldsymbol{\xi}}-\boldsymbol{\eta}}
{\sqrt{2}}\bar{\boldsymbol{\nu}}}.\nonumber \\
\end{eqnarray} 
Notice that we get an extra factor of $(-1)^{|D|}$, where $|D|$ is the number of sites in the region $D$,
which is the Jacobian for the change of variables of $\boldsymbol{\xi}$ and $\bar{\boldsymbol{\eta}}$. 
Using the formula for Gaussian integrals, we can also calculate the second integral,
\begin{eqnarray}\label{Mlambda5}\
M(\bar{\lambda}_{D})=
\det[\boldsymbol{F}]\det[\frac{1}{2}(\mathbb{1}-\boldsymbol{G}^{(ba)})]e^{-\frac{1}{2}{\rm tr\boldsymbol{Y}}} \nonumber \\  
\int d\boldsymbol{\eta}\int d{\bar{\boldsymbol{\xi}}}\,\,e^{\frac{1}{2}\bar{\boldsymbol{\xi}}\boldsymbol{X}
\bar{\boldsymbol{\xi}}+\frac{1}{2}\boldsymbol{\eta}\boldsymbol{Z}\boldsymbol{\eta}+\bar{\boldsymbol{\xi}}e^{\boldsymbol{Y}}\boldsymbol{\eta}+\frac{\bar{\boldsymbol{\xi}}+\boldsymbol{\eta}}{\sqrt{2}}\boldsymbol{F}^{-1}
\frac{\boldsymbol{\eta}-\bar{\boldsymbol{\xi}}}{\sqrt{2}}}. 
\end{eqnarray} 
The $(-1)^{|D|}$ cancels out as the $\det[\boldsymbol{F}]$ also brings down  $(-1)^{|D|}$ factor.
Furthermore, we introduce variables by defining
$\bar{\boldsymbol{\alpha}}=\frac{\bar{\boldsymbol{\xi}}+\boldsymbol{\eta}}{\sqrt{2}}$ and 
$\boldsymbol{\alpha}=\frac{-\bar{\boldsymbol{\xi}}+\boldsymbol{\eta}}{\sqrt{2}}$,
\begin{eqnarray}\label{Mlambda7}\
M(\bar{\lambda}_{D})=\det[\boldsymbol{F}]\det[\frac{1}{2}(\mathbb{1}-\boldsymbol{G}^{(ba)})]e^{-\frac{1}{2}{\rm tr \boldsymbol{Y}}}\nonumber \\ \int d\boldsymbol{\alpha}
\int d{\bar{\boldsymbol{\alpha}}}\,\,e^{\frac{1}{2}\bar{\boldsymbol{\alpha}}e^{\textbf{Y}}\bar{\boldsymbol{\alpha}}-\frac{1}{2}\boldsymbol{\alpha}e^{\textbf{Y}}\boldsymbol{\alpha}+\bar{\boldsymbol{\alpha}}[-\textbf{X}+
e^{\textbf{Y}}]\boldsymbol{\alpha}+\bar{\boldsymbol{\alpha}}\boldsymbol{F}^{-1}\boldsymbol{\alpha}}.
\end{eqnarray} 
Since $e^{\boldsymbol{Y}}$ is symmetric, the two terms $\bar{\boldsymbol{\alpha}}e^{\textbf{Y}}\bar{\boldsymbol{\alpha}}$ and $\boldsymbol{\alpha}e^{\textbf{Y}}\boldsymbol{\alpha}$  have no contribution in the integrals. Finally, after performing the integrals, we have
\begin{eqnarray}\label{Main1}\
M(\bar{\lambda}_{D})= e^{-\frac{1}{2}{\rm tr \boldsymbol{Y}}}\det[\frac{1}{2}(\mathbb{1}-\boldsymbol{G}^{(ba)})]\det[\mathbb{1}-\textbf{F}\textbf{X}+\textbf{F}e^{\textbf{Y}}].\nonumber\\.
\end{eqnarray} 
Using the definition of $\textbf{F}=(\boldsymbol{G}^{(ba)}+\mathbb{1})(\mathbb{1}-\boldsymbol{G}^{(ba)})^{-1}$, we get
\begin{eqnarray}\label{Main2}\
&&M(\bar{\lambda}_{D})=\det[e^{\textbf{Y}}]^{\frac{1}{2}}\nonumber \\ &&\det[\frac{\mathbb{1}-\boldsymbol{G}^{(ba)}}{2}
e^{-\textbf{Y}}+\frac{\mathbb{1}+\boldsymbol{G}^{(ba)}}{2}(\mathbb{1}-\textbf{X}e^{\textbf{-Y}})],
\end{eqnarray} 
where we have used the identity $[e^{-{\rm tr}\textbf{Y}}]^{\frac{1}{2}}=(\det[e^{-\textbf{Y}}])^{\frac{1}{2}}$. By introducing the definition of $\textbf{X}$ and $\textbf{Y}$ we can also write:
\begin{eqnarray}\label{Main3}\
&&M(\bar{\lambda}_{D})=
\det[\textbf{T}_{22}]^{-\frac{1}{2}}\nonumber \\ &&\det[\frac{\mathbb{1}-\boldsymbol{G}^{(ba)}}{2}\textbf{T}_{22}+\frac{\mathbb{1}+
\boldsymbol{G}^{(ba)}}{2}(\mathbb{1}-\textbf{T}_{12})].
\end{eqnarray} 
Although the above three equations are equivalent to the formula  derived in earlier works, see for example Ref.\cite{Klich2014}, they have different forms. We will use the current forms to calculate the comulants in the next sections. Note that one can consider the above equations 
as the generating function of any observable defined in
the subsystem $D$ that can be written as the quadratic equation (\ref{HD1}). Consequently, all the subsequent formulas are 
valid for any observable that can be written in a quadratic form in the fermionic representation.

\subsubsection{T matrix expansion}
One can determine the exact form of the matrix $\textbf{T}$ by expanding it with respect to ${\lambda}_{D}$ as follows:

\begin{eqnarray}\label{Texpand}\
\textbf{T}=\mathbb{1}+&&\bar{\lambda}_{D}\begin{pmatrix}
\textbf{A} & \textbf{B}\\
\textbf{-B} & \textbf{-A}\\
  \end{pmatrix}  +
  \frac{\bar{\lambda}_{D}^2}{2!}\begin{pmatrix}
\textbf{A} & \textbf{B}\\
\textbf{-B} & \textbf{-A}\\
  \end{pmatrix}  ^2+ \nonumber \\
    &&\frac{\bar{\lambda}_{D}^3}{3!}\begin{pmatrix}
\textbf{A} & \textbf{B}\\
\textbf{-B} & \textbf{-A}\\
  \end{pmatrix}  ^3+... .
\end{eqnarray} 
By calculating the right-hand side of the above equation, we can see that there are only two independent block matrices which can be obtained by using the following formulas:
\begin{eqnarray}\label{T12}\
\textbf{T}_{12}&=&\sum_{n=1}^{\infty}\frac{\bar{\lambda}_{D}^{n}}{n!}\boldsymbol{\tau}_{n}^{(1)},\\
\label{T22}
\textbf{T}_{22}&=&\mathbb{1}+\sum_{n=1}^{\infty}\frac{\bar{\lambda}_{D}^{n}}{n!}\boldsymbol{\tau}_{n}^{(2)}.
\end{eqnarray} 
Then for the other two matrices we simply have
\begin{eqnarray}\label{T11}\
\textbf{T}_{11}=\textbf{T}_{22}\,\,(\bar{\lambda}_{D}\rightarrow -\bar{\lambda}_{D}),
\end{eqnarray} 
\begin{eqnarray}\label{T21}\
\textbf{T}_{21}=\textbf{T}_{12}\,\,(\bar{\lambda}_{D}\rightarrow -\bar{\lambda}_{D}).
\end{eqnarray} 
The $\boldsymbol{\tau}_{n}^{(1,2)}$ can be calculated as follows:
\begin{eqnarray}\label{Tau1odd}\
\boldsymbol{\tau}_{odd}^{(1)}&=&\sum_{\{n_{i}\}}\, \textbf{A}^{n_{1}}\,\textbf{B}^{n_{2}}...\, 
\textbf{A}^{n_{k-1}}\,\textbf{B}^{n_{k}}sgn(k), \nonumber \\ 
&&\sum_{j=1}^{k/2}n_{2j-1}=even, \nonumber \\
\boldsymbol{\tau}_{even}^{(1)}&=&
\sum_{\{n_{i}\}}\, \textbf{A}^{n_{1}}\,\textbf{B}^{n_{2}}...\, \textbf{A}^{n_{k-1}}\,\textbf{B}^{n_{k}}sgn(k),
\nonumber \\ &&\sum_{j=1}^{k/2}n_{2j-1}=odd, 
\end{eqnarray}
and
\begin{eqnarray}\label{Tau1even}\
\boldsymbol{\tau}_{odd}^{(2)}&=&-
\sum_{\{n_{i}\}}\, \textbf{A}^{n_{1}}\,\textbf{B}^{n_{2}}...\, \textbf{A}^{n_{k-1}}\,\textbf{B}^{n_{k}}sgn(k),
\nonumber \\ &&\sum_{j=1}^{k/2}n_{2j-1}=odd, \nonumber \\
\boldsymbol{\tau}_{even}^{(2)}&=&
\sum_{\{n_{i}\}}\, \textbf{A}^{n_{1}}\,\textbf{B}^{n_{2}}...\, \textbf{A}^{n_{k-1}}\,\textbf{B}^{n_{k}}sgn(k),
\nonumber \\ &&\sum_{j=1}^{k/2}n_{2j-1}=even, 
\end{eqnarray}
where $\sum_{i=1}^{k}{n_i}=n$ and there are $2^{n-1}$ terms. Moreover, the sign of the terms can be calculated by
\begin{equation}\label{sgn}\
sgn(k)=(-1)^{\sum_{s=0(2)}^{k-2}[\frac{n_{k-s}}{2}+\frac{1-(-1)^{\sum_{j=0}^{s-1}n_{k-j}}}{4}]},
\end{equation} 
where $[x]$  is the floor (the largest integer less than or equal to $x$) and for $s<1$, we have ${\sum_{j=0}^{s-1}n_{k-j}}=0$, for more details see Appendix A.

\subsubsection{Subsystem energy moments from the generating function}
Since the exact form of $\textbf{T}_{22}$ and $\textbf{T}_{12}$ are known, one can  calculate all the moments using the equation (\ref{Main3}). To start, we need to use the expansion of
functions of determinant provided in Ref.\cite{Withers2010}. Then we have
\begin{eqnarray}\label{T22}\
\det[\textbf{T}_{22}]^{\delta}=1+\sum_{k=1}^{\infty}t_{k}\frac{\bar{\lambda}_{D}^{k}}{k!},
\end{eqnarray} 
where 
\begin{eqnarray}\label{T222}\
t_{k}=\sum_{j=1}^{k}f_{j}B_{kj}(g);
\end{eqnarray} 
and $f_{j}=\delta^{j}$ and  $B_{kj}(g)$ is the partial exponential Bell's polynomial (see Appendix B) defined as
\begin{eqnarray}\label{Bell polynomial}
\frac{1}{j!}\Big{(}\sum_{k=1}^{\infty}\frac{g_k\epsilon^k}{k!}\Big{)}^j=\sum_{k=j}^{\infty}B_{kj}(g)\frac{\epsilon^k}{k!}.
\end{eqnarray}
Here, we list the first few terms,
\begin{subequations}
\begin{eqnarray}\label{E with respect to g}
t_0&=&1,\\
t_1&=&f_1g_1, \\
t_2&=&f_1g_2+f_2g_1^2,\\
t_3&=&f_1g_3+f_2(3g_1g_2)+f_3g_1^3,\\
t_4&=&f_1g_4+f_2(4g_1g_3+3g_2^2)\nonumber \\
&+&f_3(6g_1^2g_2)+f_4g_1^4,
\end{eqnarray}
\end{subequations}
and $g=(g_1,g_2,...)$ with
\begin{eqnarray}\label{g}
g_k=\sum_{j=1}^k(-1)^{j-1}(j-1)!{\rm tr} B_{kj}(\boldsymbol{\tau}^{(2)}),
\end{eqnarray}
where $\boldsymbol{\tau}^{(2)}=(\boldsymbol{\tau}_1^{(2)},\boldsymbol{\tau}_2^{(2)},...)$. The ${\rm tr} B_{kj}(\boldsymbol{\tau}^{(2)})$ can be evaluated by calculating $ B_{kj}(g)$ and then symmetrization of all the terms 
$G_1G_2..G_r\to  \frac{1}{r!}\sum_r G_{\pi_1}G_{\pi_2}...G_{\pi_r}$, where the $G_i$'s are any sequence of the $\{g_k\}$ and the sum is over all permutations.
After having a symmetrized form for $ B_{kj}(g)$, we can now replace $\{g_k\}$ with $\{\boldsymbol{\tau}^{(2)}_k\}$ and derive the formulas for ${\rm tr} B_{kj}(\boldsymbol{\tau}^{(2)})$. Here, we list a few of the coefficients,
\begin{subequations}
\begin{eqnarray}\label{g with respect to tau}
g_1&=&{\rm tr} \boldsymbol{\tau}_1^{(2)},\\
g_2&=&{\rm tr}[\boldsymbol{\tau}_2^{(2)}-(\boldsymbol{\tau}_1^{(2)})^2],\\
g_3&=&{\rm tr}[\boldsymbol{\tau}_3^{(2)}-3\boldsymbol{\tau}_1^{(2)}\boldsymbol{\tau}_2^{(2)}+2(\boldsymbol{\tau}_1^{(2)})^3],\\
g_4&=&{\rm tr}[\boldsymbol{\tau}_4^{(2)}-4\boldsymbol{\tau}_1^{(2)}\boldsymbol{\tau}_3^{(2)}-3(\boldsymbol{\tau}_2^{(2)})^2\nonumber \\&+&12(\boldsymbol{\tau}_1^{(2)})^2\boldsymbol{\tau}_2^{(2)}-6(\boldsymbol{\tau}_1^{(2)})^4].
\end{eqnarray}
\end{subequations}
Note that in our case, $\delta=-\frac{1}{2}$.  We need to also calculate the second determinant in the equation (\ref{Main3}). Inside the determinant can be written as:
\begin{eqnarray}\label{T21t12}
&&\frac{\mathbb{1}-\boldsymbol{G}^{(ba)}}{2}\textbf{T}_{22}+\frac{\mathbb{1}+\boldsymbol{G}^{(ba)}}{2}(\mathbb{1}-\textbf{T}_{12})=\nonumber \\
&&\frac{\mathbb{1}-\boldsymbol{G}^{(ba)}}{2}(\mathbb{1}+\sum_{n=1}^{\infty}\frac{(\bar{\lambda})^{n}}{n!}\boldsymbol{\tau}_{n}^{(2)})+\nonumber \\ &&\frac{\mathbb{1}+\boldsymbol{G}^{(ba)}}{2}(\mathbb{1}-\sum_{n=1}^{\infty}\frac{\bar{\lambda}^{n}}{n!}
\boldsymbol{\tau}_{n}^{(1)}) \nonumber \\
&=&\mathbb{1}+\sum_{n=1}^{\infty}\frac{\bar{\lambda}^{n}}{n!}[\frac{\mathbb{1}-\boldsymbol{G}^{(ba)}}{2}\boldsymbol{\tau}_n^{(2)}-\frac{\mathbb{1}+\boldsymbol{G}^{(ba)}}{2}\boldsymbol{\tau}_n^{(1)}].
\end{eqnarray}
Then, we define the following matrices:
\begin{eqnarray}\label{taun}\
\tilde{\boldsymbol{\tau}}_n=
\frac{\mathbb{1}-\boldsymbol{G}^{(ba)}}{2}\boldsymbol{\tau}_n^{(2)}-\frac{\mathbb{1}+\boldsymbol{G}^{(ba)}}{2}\boldsymbol{\tau}_n^{(1)}.
\end{eqnarray} 
We can now calculate $\det[\mathbb{1}+\sum_{n=1}^{\infty}\frac{\bar{\lambda}^{n}}{n!}\tilde{\boldsymbol{\tau}}_n]$ as before with the condition $\delta=1$ which implies $f_{j}=1$.
The expansion has the following form:
\begin{eqnarray}\label{Mlambda10}\
\det[\mathbb{1}+
\sum_{n=1}^{\infty}\frac{\bar{\lambda}_{l}^{n}}{n!}\tilde{\boldsymbol{\tau}}_n]=\mathbb{1}+
\sum_{k=1}^{\infty}\tilde{t}_{k}\frac{\bar{\lambda}_{D}^{k}}{k!},
\end{eqnarray} 
where
\begin{eqnarray}\label{T222}\
\tilde{t}_{k}=\sum_{j=1}^{k}B_{kj}(\tilde{g}).
\end{eqnarray} 
The first few terms have the following forms:
\begin{subequations}
\begin{eqnarray}\label{E with respect to gtilde}
\tilde{t}_0&=&1,   \\ 
\tilde{t}_1&=&\tilde{g}_1,   \\ 
\tilde{t}_2&=&\tilde{g}_2+\tilde{g}_1^2, \\
\tilde{t}_3&=&\tilde{g}_3+3\tilde{g}_1\tilde{g}_2+\tilde{g}_1^3,\\
\tilde{t}_4&=&\tilde{g}_4+4\tilde{g}_1\tilde{g}_3+3\tilde{g}_2^2+6\tilde{g}_1^2\tilde{g}_2+\tilde{g}_1^4,
\end{eqnarray}
\end{subequations}
with
\begin{subequations}
\begin{eqnarray}\label{g with respect to tautilde}
\tilde{g}_1&=&{\rm tr} \tilde{\boldsymbol{\tau}}_1,\\
\tilde{g}_2&=&{\rm tr}[\tilde{\boldsymbol{\tau}}_2-\tilde{\boldsymbol{\tau}}_1^2],\\
\tilde{g}_3&=&{\rm tr}[\tilde{\boldsymbol{\tau}}_3-3\tilde{\boldsymbol{\tau}}_1\tilde{\boldsymbol{\tau}}_2+2(\tilde{\boldsymbol{\tau}}_1)^3],\\
\tilde{g}_4&=&{\rm tr}[\tilde{\boldsymbol{\tau}}_4-4\tilde{\boldsymbol{\tau}}_1\tilde{\boldsymbol{\tau}}_3-3(\tilde{\boldsymbol{\tau}}_2)^2+12(\tilde{\boldsymbol{\tau}}_1)^2\tilde{\boldsymbol{\tau}}_2\nonumber \\&-&6(\tilde{\boldsymbol{\tau}}_1)^4].
\end{eqnarray}
\end{subequations}
Using the two formulas (\ref{T22}) and (\ref{Mlambda10}) the expansion for $M(\bar{\lambda}_{D})$ becomes 
\begin{eqnarray}\label{Mlambda11}
M(\bar{\lambda}_{D})&=&(\sum_{k=0}^{\infty}t_{k}\frac{\bar{\lambda}_{D}^k}{k!})(\sum_{k'=0}^{\infty}\tilde{t}_{k'}\frac{\bar{\lambda}_{D}^{k'}}{k'!}) \nonumber \\
&=&\sum_{m=0}^{\infty}\sum_{j=0}^{m}t_{j}\tilde{t}_{m-j}\frac{\bar{\lambda}_{D}^j}{j!}\frac{\bar{\lambda}_{D}^{m-j}}{(m-j)!} \nonumber \\
&=&\sum_{m=0}^{\infty}\sum_{j=0}^{m}\frac{\bar{\lambda}_{D}^m}{j!(m-j)!}t_{j}\tilde{t}_{m-j},
\end{eqnarray}
which simplifies to
\begin{eqnarray}\label{Mlambda11}
M(\bar{\lambda}_{D})=\sum_{m=0}^{\infty}\frac{\bar{\lambda}_{D}^m}{m!}\sum_{j=0}^{m}{{m}\choose{j}}t_{j}\tilde{t}_{m-j}.
\end{eqnarray}
An expansion of the full counting statistics with respect to $\lambda$ can be written as below:
\begin{eqnarray}\label{determinant expansion}
M(\bar{\lambda}_{D})=1+E_1\bar{\lambda}_{D}+E_2\frac{\bar{\lambda}_{D}^2}{2!}+...,
\end{eqnarray}
where we have $E_m=\langle  H_D^m\rangle$. Using this definition, one can calculate all the moments as,
\begin{eqnarray}\label{Em general}
E_m=\langle H_{D}^m\rangle =\sum_{j=0}^{m}{{m}\choose{j}}t_{j}\tilde{t}_{m-j}.
\end{eqnarray}
One can  directly check that the above equation for $m=1$ and $2$ produces the equations (\ref{Hl3}) and (\ref{Hl6}), respectively.
For the future discussion, it is also important to introduce 
 the fluctuation of the $m$ moment  defined as:
 \begin{eqnarray}\label{fluctuation m moment}
\tilde{E}_{m}=\langle (H_{D}-E_1)^{m}\rangle.
\end{eqnarray}
Consequently, by using the equation (\ref{Em general}) and working out some algebra, one can show that
\begin{subequations}
\begin{eqnarray}\label{Hl1}
\tilde{E}_1&=& 0, \\
\tilde{E}_2&=& \tilde{g}_2-\frac{g_2}{2}, \\
\tilde{E}_3&=& \tilde{g}_3-\frac{g_3}{2}, \\
\tilde{E}_4-3\tilde{E}_2^2&=& \tilde{g}_4-\frac{g_4}{2}, \\
\tilde{E}_5-10\tilde{E}_2\tilde{E}_3&=& \tilde{g}_5-\frac{g_5}{2}.
\end{eqnarray}
\end{subequations}
Finally, the most general case can be written as:
\begin{eqnarray}\label{ E tilde general}
\tilde{E}_m&+&\sum_{j=2}^{m}(-1)^{j-1}(j-1)!B_{mj}(\tilde{E}_1,\tilde{E}_2,...,\tilde{E}_{m-j+1})\nonumber\\
&=&\tilde{g}_m-\frac{g_m}{2},\hspace{1cm}m>1
\end{eqnarray}
where the $B_{mj}$ is the Bell's polynomial as we mentioned before. 
 The above formulas show that the quantities $g_i$ and $\tilde{g}_i$ with $i>2$ represent the fluctuations of the subsystem energy.
 
\subsection{Subsystem energy  generating function for an arbitrary state}
In this section, we study the subsystem energy  generating function for an arbitrary state. In other words, we would like to calculate 
\begin{eqnarray}\label{Mlambda1 arbitrary state}
M(\bar{\lambda}_{D})=
\langle\psi| e^{\bar{\lambda}_{D}H_{D}}|\psi\rangle,
\end{eqnarray} 
 for arbitrary state $|\psi\rangle$. 
 To calculate the above quantity, we assume that we know the form of $|\psi\rangle$ in the fermion occupation basis, in other words, we have
 \begin{eqnarray}\label{ arbitrary state}
|\psi\rangle=\sum_{\{ C \}}a_C|C\rangle,
\end{eqnarray} 
 where $|C\rangle$ is an arbitrary configuration for fermions in the subsystem. Using the above state in the equation(\ref{Mlambda1 arbitrary state}), we have
 \begin{eqnarray}\label{Mlambda1 arbitrary state2}
M(\bar{\lambda}_{D})=\sum_{\{ C' \}}\sum_{\{ C \}}a^*_{C'}a_C
\langle C'| e^{\bar{\lambda}_{D}H_{D}}|C\rangle.
\end{eqnarray} 
The quantity $\langle C'| e^{\bar{\lambda}_{D}H_{D}}|C\rangle$ can be calculated using the equation (\ref{e_lambda_l}) 
as follows (see Ref.\cite{Najafi2016}): Consider an arbitrary configuration $C$. Then, the corresponding Grassmann variable for the unoccupied state is zero while for occupied state, one needs to integrate over Grassmann variable. This transformation will effectively create new matrices $\boldsymbol{X}_{C'}$, $\boldsymbol{Z}_C$ and $(e^{\boldsymbol{Y}})_{C'C}$ that are dependent on the configurations. Then we can write
\begin{eqnarray}\label{e_lambda_l2}
M(\bar{\lambda}_{D})=\sum_{\{ C' \}}\sum_{\{ C \}}a^*_{C'}a_C\int\prod_{C'}d\bar{\xi}\int\prod_{C}d\bar{\eta}
\nonumber \\e^{\frac{1}{2}\bar{\boldsymbol{\xi}}\boldsymbol{X}_{C'}\bar{\boldsymbol{\xi}}+\frac{1}{2}\boldsymbol{\eta}\boldsymbol{Z}_{C}\boldsymbol{\eta}-
\frac{1}{2}{\rm tr \boldsymbol{Y}}+\bar{\boldsymbol{\xi}}(e^{\boldsymbol{Y}})_{C'C}\boldsymbol{\eta}}.
\end{eqnarray}
Finally, after performing the Grassmann integration, we have
 \begin{eqnarray}\label{final}
M(\bar{\lambda}_{D})=e^{-\frac{1}{2}\text{tr}\boldsymbol{Y}}\sum_{\{ C' \}}\sum_{\{ C \}}a^*_{C'}a_C\text{pf}\begin{bmatrix}
       \boldsymbol{X}_{C'} & (e^{\boldsymbol{Y}})_{C'C}          \\[0.3em]
       -(e^{\boldsymbol{Y}})_{CC'}^T & \boldsymbol{Z}_{C}          
     \end{bmatrix},\nonumber \\
\end{eqnarray}
where $\text{pf}$ is the Pfaffian of the matrix. The above equation can be very useful for those states that have a simple form in the configuration basis.
It can be also useful in the study of time dependent systems.
For example, for the state without any fermion, we simply have
\begin{eqnarray}\label{e_lambda_l no fermion}
M(\bar{\lambda}_{D})=e^{-\frac{1}{2}\text{tr}\boldsymbol{Y}},
\end{eqnarray}
and for the case with full of fermions, we have
\begin{eqnarray}\label{e_lambda_l full fermion}
M(\bar{\lambda}_{D})=e^{-\frac{1}{2}\text{tr}\boldsymbol{Y}}\text{pf}\begin{bmatrix}
       \boldsymbol{X} & e^{\boldsymbol{Y}}          \\[0.3em]
       -(e^{\boldsymbol{Y}})^T & \boldsymbol{Z}         
     \end{bmatrix}.
\end{eqnarray}
%

\section{ Subsystem energy statistics in the $XY$ spin chain}

In this section, we use the equations that we derived in the previous section to study the full counting statistics of the subsystem energy for the $XY$ spin chain. We study different phases of the chain with analytical
and numerical techniques.
In particular, we study the transverse field Ising chain and XX chain in more details.

\subsection{Definitions and general results}
\begin{figure} [htbp] 
\includegraphics[width=0.45\textwidth,angle =0]{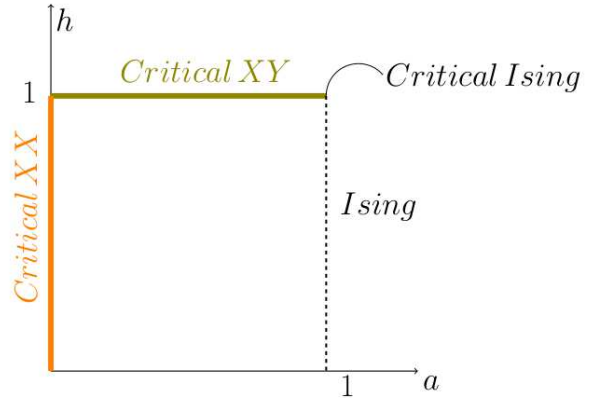}
\caption{Different critical regions in the quantum $XY$ chain. The critical $XX$ 
chain has central charge $c=1$ and critical $XY$ chain has $c=\frac{1}{2}$. } 
\label{fig:XYpahse space}
\end{figure}

The Hamiltonian of the  $XY$-chain is as follows
\begin{eqnarray}\label{HXY1}\
H_{XY}=&-&\frac{J}{2}\sum_{j=1}^{L}\Big{[}(\frac{1+a}{2})\sigma_{j}^{x}\sigma_{j+1}^{x}+(\frac{1-a}{2})\sigma_{j}^{y}\sigma_{j+1}^{y}\Big{]}\nonumber \\&-&\frac{h}{2}\sum_{j=1}^{L}\sigma_{j}^{z},
\end{eqnarray} 
where $\sigma_{j}^{x,y,z}$  are the Pauli matrices. Here, $a$ indicates the  anisotropy interaction between spins and $h$ denotes  the transverse magnetic field.
Using the Jordan-Wigner transformation
$c_j^{\dagger}=\prod_{l<j}\sigma_l^z\sigma_j^{+}$, one can map the Hilbert space of a quantum chain of a spin $1/2$ into the Fock space of spinless fermions. Then, the  Hamiltonian becomes 
\begin{eqnarray}\label{XY-ff}
H=&&\frac{J}{2}\sum_{j=1}^{L-1} (c_j^{\dagger}c_{j+1}+ac_j^{\dagger}c_{j+1}^{\dagger}+h.c.)-\sum_{j=1}^{L}h(c_j^{\dagger}c_j-\frac{1}{2})\nonumber\\&+&\frac{J\mathcal{N}}{2}(c_L^{\dagger}c_1+ac_L^{\dagger}c_1^{\dagger}+h.c.),
\end{eqnarray}
where $c_{L+1}^{\dagger}=0$ and $c_{L+1}^{\dagger}=\mathcal{N}c_{1}^{\dagger}$ for open and periodic boundary conditions respectively with $\mathcal{N}=\prod_{j=1}^{L}\sigma_j^z=\pm1$. 
The phase diagram of the XY chain is shown in Fig~\ref{fig:XYpahse space}.

%

We are interested in studying the Hamiltonian truncated to a subsystem of  an infinite quantum chain described by 
\begin{equation}\label{Hl1}
H_{l}=\frac{J}{2}\sum_{j=1}^{l-1} (c_j^{\dagger}c_{j+1}+ac_j^{\dagger}c_{j+1}^{\dagger}+h.c.)-\sum_{j=1}^{l}h(c_j^{\dagger}c_j-\frac{1}{2}),
\end{equation}
where $l$ denotes the size of the subsystem. Note that here, we have an open system with natural boundary conditions coming from truncating the Hamiltonian. 
%
%
It is easy to see that the Hamiltonian (\ref{Hl1}) can be written in the following form,
\begin{eqnarray}\label{Hl2}\
H_{l}=\textbf{c}^{\dagger}.\textbf{A}.\textbf{c}+\frac{1}{2}\textbf{c}^{\dagger}.\textbf{B}.\textbf{c}^{\dagger}+\frac{1}{2}\textbf{c}.\textbf{B}^{T}.\textbf{c}-\frac{1}{2}{\rm tr}{\textbf{A}},
\end{eqnarray} 
 where the matrices $\textbf{A}$ and $\textbf{B}$ are:
\begin{eqnarray}\label{mat_A}\
\textbf{A}= \begin{pmatrix}
     -h     & \frac{J}{2}     & 0      &\dots           &0 \\
     \frac{J}{2}        &-h    &\frac{J}{2}        &0               &0           \\  
     0       & \frac{J}{2}      &-h     &\frac{J}{2}              &0           \\       
    \ddots      &\ddots      & \ddots &\ddots    &\ddots               \\  
0  &0            &\dots &\frac{J}{2}  & -h  \\ 
  \end{pmatrix}  \nonumber, 
\end{eqnarray} 
\begin{eqnarray}\label{mat_B}\
\textbf{B}= \begin{pmatrix}
     0       & \frac{Ja}{2}     & 0      &\dots        &0\\
     -\frac{Ja}{2}     &0      &\frac{Ja}{2}          & &0                \\  
     0       & -\frac{Ja}{2}    &0    &\frac{Ja}{2}    & 0                 \\       
     \ddots       &\ddots      & \ddots &\ddots    &\ddots           \\ 
0 &0   & \dots&-\frac{Ja}{2} &0 \\ 
  \end{pmatrix}.
\end{eqnarray} 

Using the results of the previous section and considering $J=1$, first one can calculate $\langle H_{l}\rangle$ as follows:

\begin{eqnarray}\label{Dnm}\
\langle H_{l}\rangle =&&=\frac{1}{4}{\rm tr}[\textbf{D}^{T}\boldsymbol{G}^{(ba)}]=\frac{1}{4}[-2h\delta_{nm}+(1+a)\delta_{n,m-1}\nonumber\\ &&+(1-a)\delta_{n,m+1}]\boldsymbol{G}^{(ba)}_{nm}\nonumber \\
=&&\frac{1}{4}[-2h\boldsymbol{G}^{(ba)}_{0}+(1+a)\boldsymbol{G}^{(ba)}_{-}\nonumber \\+&&(1-a)\boldsymbol{G}^{(ba)}_{+}]l-(1+a)\boldsymbol{G}^{(ba)}_{-}\nonumber\\
&&-(1-a)\boldsymbol{G}^{(ba)}_{+}.
\end{eqnarray} 
Notice that in this case $\boldsymbol{G}^{(aa)}=-\boldsymbol{G}^{(bb)}=\mathbb{1}$. Not surprisingly the $\langle H_{l}\rangle $ is proportional to the size of the subsystem.
We can also calculate the second moment as follows:
\begin{eqnarray}\label{Hl2}\
\langle H_{l}^{2}\rangle=&&\langle H_{l}\rangle^{2}+\frac{1}{16}({\rm tr}[\textbf{D}\textbf{D}^{T}]
\nonumber \\&&-{\rm tr}[\textbf{D}^{T}\boldsymbol{G}^{(ba)}\textbf{D}^{T}\boldsymbol{G}^{(ba)}]),
\end{eqnarray} 
where we have ${\rm tr}[\textbf{D}\textbf{D}^{T}]=(2J(1-a^2)+4h^2)l-2J(1-a^2)$. It is easy to see that here, $\langle H_{l}^{2}\rangle$ is proportional to $l^2$. By similar calculation, one can easily show that
\begin{eqnarray}\label{Hl general}\
\langle H_{l}^{n}\rangle \sim l^n.
\end{eqnarray} 
All of the $\boldsymbol{G}^{(ba)}$ matrices are known for the XY chain. For example, take a periodic system and then consider
the thermodynamic limit; then we have
\begin{eqnarray}\label{G XY}\
\boldsymbol{G}^{(ba)}_{nm}=-\frac{1}{2\pi}\int_0^{2\pi} dx e^{i(n-m)x}\frac{\cos[x]+i a\sin[x]-h}{\sqrt{(\cos[x]-h)^2+a^2\sin^2[x]}}.
\end{eqnarray} 
At the critical Ising point ($a=h=1$), the above equation takes the following simple form
\begin{eqnarray}\label{G critical Ising}\
\boldsymbol{G}^{(ba)}_{nm}=-\frac{1}{\pi}\frac{1}{n-m+1/2}.
\end{eqnarray} 
Then, one can easily study the $M (\bar{\lambda}_l)$  for different values of $a$ and $h$ using the equation (\ref{Main3}). 

\subsection{Conformal field theory expectations}

At the critical points, it is expected that the system can be described by conformal field theory. In particular, on the XY critical line
the system can be described by Ising field theory with central charge $c=\frac{1}{2}$ and on the XX critical line with the central charge $c=1$.
Since at the critical points we have
\begin{eqnarray}\label{H CFT}\
H^{CFT}_l=\int_0^l dx T_{00},
\end{eqnarray} 
where $T$ is the energy-momentum tensor with the scaling dimension $2$. Notice that since 
in CFT the one point function of the energy-momentum tensor is by definition zero, we need to work with $\tilde{E}$ defined in equation (\ref{Hl1}) rather than $E$.
Then, by simple dimensional analysis we expect
\begin{eqnarray}\label{moment CFT}\
\tilde{E}_n=\langle (H^{CFT}_l)^n\rangle\sim c_0+\frac{c_{-n}}{l^n},
\end{eqnarray} 
where $c_0$ is a $n$ dependent constant which is also dependent on the cut-off as $\frac{1}{a^n}$ and $c_{-n}$ is another constant. The above equation is expected to be valid at
the critical points and we will show its validity for the critical Ising point through numerical calculation and analytically for the XX chain. 

At noncritical points, it is natural to expect
\begin{eqnarray}\label{moment non-CFT}\
\tilde{E}_n\sim a_n+b_ne^{-\alpha_n l},
\end{eqnarray}
where $a_n$, $b_n$, and $\alpha_n$ are all cut-off dependent constants.
We will show the validity of the above result by numerical calculations in the next subsections. 


\subsection{ Transverse field Ising  chain}
In this section, we study the statistics of the subsystem energy in the transverse field Ising chain. In Figs. 2 and 3,  $M(\bar{\lambda})$ has shown
for different values of $l$ and the transverse magnetic field $h$. A few comments are in order: First of
all, there is no particular difference in the shape of the $M(\bar{\lambda})$ at and outside of the critical point. Secondly,
one can see that for the finite values of $l$, $M(\bar{\lambda})$ for large positive and negative values of $\bar{\lambda}$
diverges exponentially. However, the interesting point is that since the coefficient of the exponential for positive values of $\bar{\lambda}$
is very small, one can see the effect of the exponential for just relatively large values of $\bar{\lambda}$. It seems that the coefficient of the positive exponential
decreases like an exponential with respect to the size of the subsystem $l$ which makes $M(\bar{\lambda})\simeq e^{-\bar{\lambda}}$ to be a very good approximation
for a large interval of $\bar{\lambda}$. Of course, for a very large $l$ which is comparable with the system size, one does not actually expect any fluctuation
in the energy of the system which is in its ground state. That is why in the limit of large $l$, we have exactly 
\begin{eqnarray}\label{M lambda bar}\
M(\bar{\lambda})=e^{-\lambda}.
\end{eqnarray}
From Fig. 3, one can observe that even for relatively small subsystem sizes the graphs can be described perfectly by the above equation.
 \begin{figure} [htb] \label{fig2}
\includegraphics[width=0.45\textwidth,angle =0]{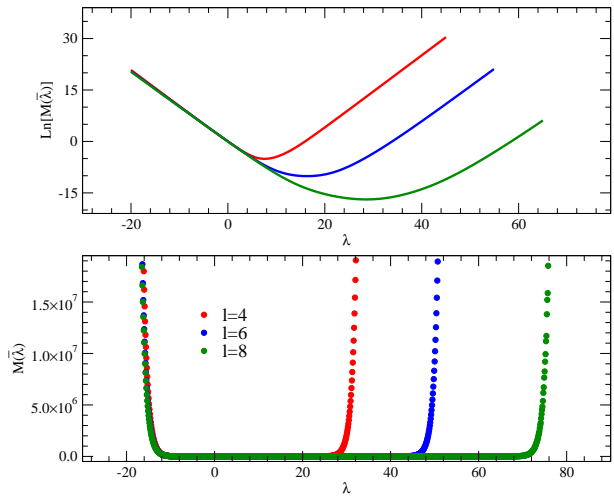}
\caption{Generating function of the subsystem energy for transverse field Ising chain with $a=h=1$ with respect to $\lambda$ for different $l$'s.} 
\end{figure}
 \begin{figure} [htb] \label{fig3}
\includegraphics[width=0.45\textwidth,angle =0]{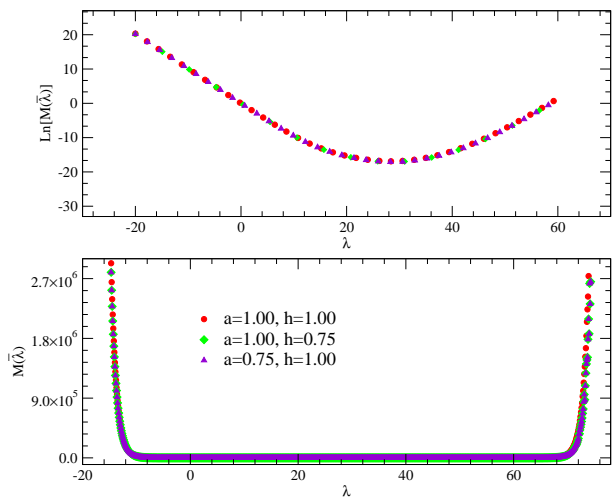}
\caption{ Generating function of the subsystem energy for XY chain  with respect to $\lambda$ for  $l=8$.}
\end{figure}
In Fig. 4, we checked the validity of the equation (\ref{moment CFT}) for $\tilde{E}_n$ with $n=2$, and $3$ at the critical point of
the transverse field Ising chain, i.e, $h=1$. The depicted numerical results confirm the CFT predictions nicely. Note that since equation (\ref{G XY})
is the thermodynamic limit of a periodic system while, the thermodynamic limit of the subsystem is an open system, one naturally does not expect $\tilde{E}_n$ to
go to zero for $l\to\infty$. In fact, It approaches a $n$-dependent constant which shows that the omitted boundary point that connects the two extremes of
the subsystem plays a finite role. Since we are just interested in the decay with respect to the subsystem size, this constant does not play any important role in our current investigation.
Outside of the critical regime,
as it is depicted in Fig. 5, the $\tilde{E}_n$ decays exponentially with respect 
to the size of the subsystem $l$, which is consistent with our prediction  in equation (\ref{moment non-CFT}).

%
 \begin{figure} [htb] \label{fig4}
\includegraphics[width=0.45\textwidth,angle =0]{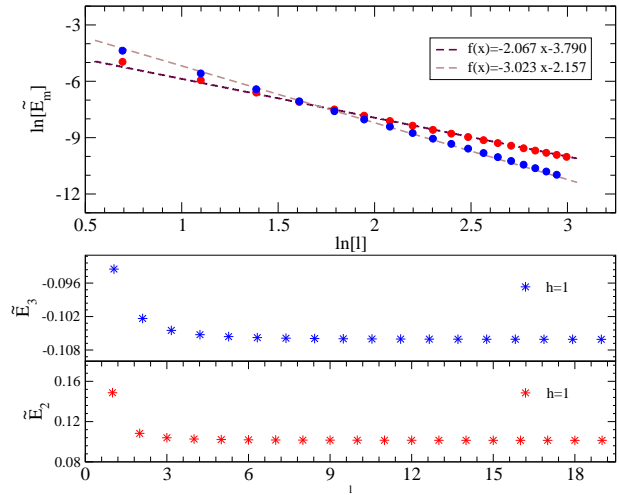}
\caption{ $\tilde{E}_2$ and $\tilde{E}_3$ with respect to the subsystem size $l$ at the critical point of
the transverse field Ising chain.} 
\end{figure}
 \begin{figure} [htb] \label{fig5}
\includegraphics[width=0.45\textwidth,angle =0]{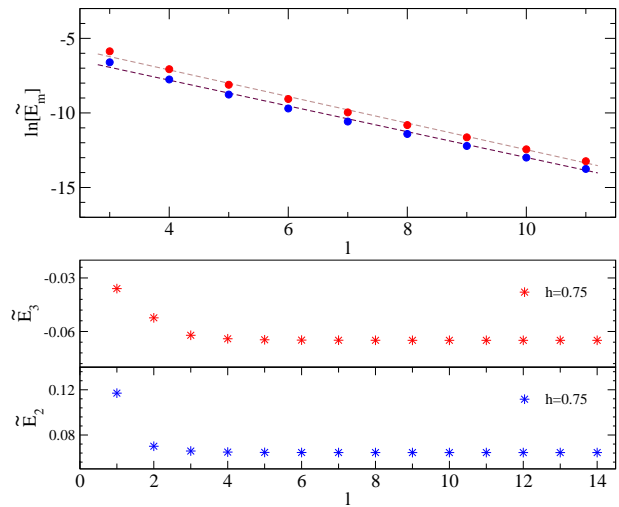}
\caption{ $\tilde{E}_2$ and $\tilde{E}_3$ with respect to the subsystem size $l$ for the noncritical 
 transverse field Ising chain with $h=0.75$.} 
\end{figure}
%

\subsection{ $XX$ spin chain}
The critical XX line is particularly interesting because it follows a simple tight binding form.
In the free fermion representation, one can calculate many quantities exactly. For this reason, we make a through discussion of
the full counting statistics in this case. We give an independent derivation of the generating function in this case which is in full agreement with the results of the previous
sections.
\subsubsection{The generating function}
The Hamiltonian of the $XX$ chain or chain of free fermions is:
\begin{eqnarray}\label{Hamiltonian}
H=-\sum_i (c_i^{\dagger}c_{i+1}+c_{i+1}^{\dagger}c_i-2\cos n_c c_i^{\dagger}c_i),
\end{eqnarray}
where $n_c$ plays the role of filling factor. We would like to calculate:
\begin{eqnarray}\label{cumulant}
M(\bar{\lambda})=\langle e^{\bar{\lambda} H_l}\rangle={\rm tr} \rho_l e^{\bar{\lambda} H_l},
\end{eqnarray}
where $H_l=-\sum_{i=1}^{l-1} (c_i^{\dagger}c_{i+1}+c_{i+1}^{\dagger}c_i-2\cos n_c c_i^{\dagger}c_i)=\sum_{ij} A_{ij}c_i^{\dagger}c_j$
with $-A_{ij}=\delta_{i,j+1}+\delta_{i+1,j}-2\cos n_c\delta_{i,j}$
and $\rho_l$ is the reduced density matrix given by Ref.\cite{Peschela},
\begin{eqnarray}\label{redused density matrix}
\rho_l=\det (1-C)e^{(\ln F)_{ij}c_i^{\dagger}c_j},
\end{eqnarray}
where $F=C(1-C)^{-1}$. For an infinite system, $C_{ij}=\langle c_i^{\dagger}c_j\rangle=\frac{\sin [n_c(i-j)]}{\pi(i-j)}$
and $C_{ii}=\frac{n_c}{\pi}$, where we will mostly work with $n_c=\frac{\pi}{2}$.
Then, we need to calculate the following trace
\begin{eqnarray}\label{trace}
M(\bar{\lambda})=\det (1-C) {\rm tr} \Big{[} e^{(\ln F)_{ij}c_i^{\dagger}c_j}e^{\bar{\lambda} A_{ij}c_i^{\dagger}c_j}\Big{]}.
\end{eqnarray}
The trace can be easily calculated 
\begin{eqnarray}\label{trace2}
M(\bar{\lambda})&=&\det (1-C) \det [1+e^{\ln F}e^{\bar{\lambda} A}]\nonumber \\&=&\det (1-C+C e^{\bar{\lambda} A}).
\end{eqnarray}
The above equation is consistent with the equation (\ref{Main3}), which provides another check for our main formula.

The above determinant does not have a simple Toeplitz form which makes further analytic calculations nontrivial. However, it can be simplified further by diagonalizing the matrix $A$
as $A=VDV^{T}$ where:
\begin{eqnarray}\label{A diagonalization}
V_{ij}&=&\sqrt{\frac{2}{l+1}}\sin[\frac{\pi i j}{l+1}],\\
-D_k&=&2\cos[\frac{\pi k}{l+1}]-2\cos n_c,
\end{eqnarray}
where $i,j,k=1,...,l$. Then, by defining $\tilde{C}=V^{T}CV$ we can write
\begin{eqnarray}\label{GF2}
M(\bar{\lambda})=\det (1-\tilde{C}+\tilde{C} e^{\bar{\lambda} D}).
\end{eqnarray}
Note that we have $V^{T}V=I$. For a reason that will be clear soon, it is better to write the above equation as
\begin{eqnarray}\label{GF2}
M(\bar{\lambda})=\det \tilde{C}\det (\tilde{F}^{-1}+ e^{\bar{\lambda} D}),
\end{eqnarray}
where $\tilde{F}^{-1}=(1-\tilde{C})/\tilde{C}$. There is an interesting formula for the determinant of the sum of an arbitrary matrix $X$ of size $n$ and a diagonal matrix $Y$ 
with elements $y_i$, $i=1,...,n$ as follows \cite{Prells2003}:
\begin{eqnarray}\label{formula for the determinant of a sum}
\det(X+Y)=\det X+y_1y_2...y_n+\sum_{i=1}^{n-1}\alpha_i\beta_i;
\end{eqnarray}
where the vectors $\alpha_i$ and $\beta_i$ with the sizes $n_i=\binom{n}{i}$ are defined as
\begin{eqnarray}\label{alpha}
\alpha_i&=&(y_{n-i+1}..y_n,...,y_1..y_i),\\
\label{beta}
\beta_i&=&([\mathcal{C}_i(X)]_{11},...,[\mathcal{C}_i(X)]_{n_in_i}),
\end{eqnarray}
where $\mathcal{C}_i(X)$ is the compound matrix of rank $i$ of the matrix $X$. It is a matrix with the  size $n_i$
formed from the determinants of all $i\times i$ submatrices 
of $X$, i.e., all  $i\times i$ minors, arranged with the submatrix index sets in lexicographic order. Note that the generic element of the vector
$\alpha_i$ can be written as $(\alpha_i)_k=(E_i\mathcal{C}_i(Y)E_i)_{kk}$, where  $\mathcal{C}_i(Y)$ is the compound matrix of rank $i$ of the matrix $Y$,
and $E_i$ is the rotated identity matrix of size $n_i$ with elements $(E_i)_{kk'}=\delta_{k,n_i-k'+1}$. Using the above result we have
\begin{widetext}
\begin{eqnarray}\label{GF numerics}
M(\bar{\lambda})=\det \tilde{C}\Big{(}\det \tilde{F}^{-1}+1+\sum_{i=1}^{l-1}\alpha_i\beta_i\Big{)}=\det(1-C)+\det C+\det C\sum_{i=1}^{l-1}\alpha_i\beta_i,
\end{eqnarray}
\end{widetext}
where the vectors $\alpha_i$ and $\beta_i$ with the sizes $l_i=\binom{l}{i}$ are defined by (\ref{alpha}) and (\ref{beta})
with $y_s=e^{-2\bar{\lambda}\cos[\frac{\pi s}{l+1}]+2\bar{\lambda}\cos n_c}$ where $s=1,2,...,l$ and $X=\tilde{F}^{-1}$. Note that it is easy to see that the possible energies are 
either $-2\cos[\frac{\pi s}{l+1}]+2\cos n_c$ or different possible summations among them.
Although the above equation is useful for numerical reasons, it can not be easily  used to calculate 
the asymptotic values of the moments. However, the formula is useful if one is interested in calculating the probability of finding the system in a single eigenstate of the Hamiltonian truncated to a subsystem. Here, we report the generating function for half filling for small subsystem sizes:
\begin{widetext}
\begin{eqnarray}\label{small l1}
M(\bar{\lambda})=\left(\frac{1}{2}-\frac{2}{\pi^2}\right)+\left(\frac{1}{4}+\frac{1}{\pi}+\frac{1}{\pi^2}\right)e^{-\bar{\lambda}}+\left(\frac{1}{4}-\frac{1}{\pi}+\frac{1}{\pi^2}\right)e^{+\bar{\lambda}}\hspace{5cm}l=2
\end{eqnarray}
\begin{eqnarray}\label{small l2}  
M(\bar{\lambda})=\left(\frac{1}{2}-\frac{4}{\pi ^2}\right)+\left(\frac{1}{4}+\frac{2}{\pi ^2}+\frac{\sqrt{2}}{\pi }\right) e^{-\sqrt{2}\,\, \bar{\lambda}
   }+\left(\frac{1}{4}+\frac{2}{\pi ^2}-\frac{\sqrt{2}}{\pi }\right) e^{\sqrt{2}\,\, \bar{\lambda} }\hspace{3.7cm}l=3
\end{eqnarray}
\begin{eqnarray}
M(\bar{\lambda})=&&\left(\frac{1}{4}+\frac{64}{9 \pi ^4}-\frac{46}{15 \pi ^2}\right)+\left(\frac{1}{16}+\frac{16}{9 \pi ^4}-\frac{8}{9 \pi ^3}-
\frac{5}{9 \pi^2}+\frac{1}{6 \pi }\right) e^{-\bar{\lambda} }+
\left(\frac{1}{16}+\frac{16}{9 \pi ^4}+\frac{8}{9 \pi ^3}-\frac{5}{9 \pi ^2}-\frac{1}{6 \pi }\right) e^{+\bar{\lambda} }\nonumber \\
   &+&\left(\frac{1}{16}+\frac{16}{9 \pi ^4}+\frac{64}{9 \sqrt{5} \pi ^3}+\frac{94}{45 \pi ^2}+\frac{4}{3 \sqrt{5} \pi }\right) e^{-\sqrt{5}\,\,\bar{\lambda} }+
   \left(\frac{1}{16}+\frac{16}{9 \pi ^4}-\frac{64}{9 \sqrt{5} \pi ^3}+\frac{94}{45 \pi ^2}-\frac{4}{3 \sqrt{5} \pi }\right) e^{\sqrt{5}\,\, \bar{\lambda}}\nonumber \\
   &+&\left(\frac{1}{8}-\frac{32}{9 \pi ^4}+\frac{8}{9 \pi ^3}-\frac{64}{9 \sqrt{5} \pi ^3}+\frac{16}{9 \sqrt{5} \pi ^2}+\frac{1}{6 \pi
   }+\frac{4}{3 \sqrt{5} \pi }\right) e^{-\frac{\sqrt{5}\,\, \bar{\lambda}}{2}-\frac{\bar{\lambda} }{2}}\nonumber \\
&+&\left(\frac{1}{8}-\frac{32}{9 \pi ^4}-\frac{8}{9 \pi ^3}+\frac{64}{9 \sqrt{5} \pi ^3}+\frac{16}{9 \sqrt{5} \pi ^2}-\frac{1}{6 \pi }-\frac{4}{3 \sqrt{5} \pi }\right) e^{\frac{\sqrt{5}\,\,
   \bar{\lambda} }{2}+\frac{\bar{\lambda} }{2}}\nonumber \\
   &+&\left(\frac{1}{8}-\frac{32}{9 \pi ^4}-\frac{8}{9 \pi ^3}-\frac{64}{9 \sqrt{5} \pi ^3}-\frac{16}{9 \sqrt{5} \pi ^2}-\frac{1}{6 \pi }+\frac{4}{3 \sqrt{5} \pi }\right)
   e^{-\frac{\sqrt{5}\,\,\bar{ \lambda} }{2}+\frac{\bar{\lambda} }{2}}\nonumber\\
   &+&\left(\frac{1}{8}-\frac{32}{9 \pi ^4}+\frac{8}{9 \pi ^3}+\frac{64}{9 \sqrt{5} \pi ^3}-\frac{16}{9 \sqrt{5} \pi ^2}+\frac{1}{6 \pi }-\frac{4}{3 \sqrt{5} \pi
   }\right) e^{\frac{\sqrt{5}\,\, \bar{\lambda} }{2}-\frac{\bar{\lambda} }{2}} \hspace{2cm}l=4  
\end{eqnarray}
\end{widetext}
The numbers in the exponentials are the possible subsystem energy values and the coefficients of the exponentials are the probability of occurrence of the corresponding subsystem energy. The coefficients are highly nontrivial numbers.
In  Fig. $6$, we depicted the distribution of the 
logarithm of the subsystem energy for the XX chain. As it is clear, although the probability decreases exponentially it is not a smooth function.
The exponential decrease of the probability just means that the subsystem with high probability is either in its ground state or in its first few excited states.
It is natural to expect that this should be true independent of the considered model. 
 \begin{figure} [htb] \label{fig6}
\includegraphics[width=0.45\textwidth,angle =0]{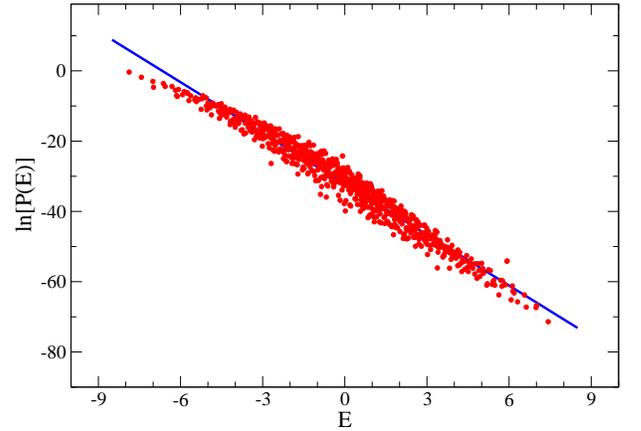}
\caption{Logarithm of the probability of finding the subsystem in different energies for the XX chain with $l=13$.} 
\end{figure}
%
 \subsubsection{Calculation of moments}
In this subsection, we calculate the subsystem energy moments in the XX chain.
To do that one can  expand the equation (\ref{trace2}) directly and derive some results for the moments as we did in the Sec. II.  First of all, we write
\begin{eqnarray}\label{determinant}
M(\bar{\lambda})=\det (1+\sum_{i=1}^{\infty}\frac{\bar{\lambda}^m}{m!}CA^m )
\end{eqnarray}
An expansion of the above formula with respect to $\bar{\lambda}$ can be derived as follows:
\begin{eqnarray}\label{determinant expansion}
M(\bar{\lambda})=1+E_1\bar{\lambda}+E_2\frac{\bar{\lambda}^2}{2!}+...,
\end{eqnarray}
where we have $E_k=\langle  H_l^k\rangle$. First, we define 
\begin{subequations}
\begin{eqnarray}\label{ a}
a_0&=&1\\
a_k&=&CA^k\,\,\,\,\,\,\,\,\,\,\, k=1,2,3,....
\end{eqnarray}
\end{subequations}
Then, using the equations of Sec. II, the first few  $E_k$'s can be written as:

\begin{subequations}
\begin{eqnarray}\label{E with respect to g}
E_1&=&\mathfrak{g}_1,\\
E_2&=&\mathfrak{g}_2+\mathfrak{g}_1^2,\\
E_3&=&\mathfrak{g}_3+3\mathfrak{g}_1\mathfrak{g}_2+\mathfrak{g}_1^3,\\
E_4&=&\mathfrak{g}_4+4\mathfrak{g}_1\mathfrak{g}_3+3\mathfrak{g}_2^2+6\mathfrak{g}_1^2\mathfrak{g}_2+\mathfrak{g}_1^4,\\
E_5&=&\mathfrak{g}_5+5\mathfrak{g}_1\mathfrak{g}_4+10\mathfrak{g}_2\mathfrak{g}_3+10\mathfrak{g}_1^2\mathfrak{g}_3\nonumber \\&&+15\mathfrak{g}_1\mathfrak{g}_2^2+10\mathfrak{g}_1^3\mathfrak{g}_2+\mathfrak{g}_1^5,
\end{eqnarray}
\end{subequations}
where
\begin{subequations}
\begin{eqnarray}\label{g with respect to a}
\mathfrak{g}_1&=&{\rm tr} a_1,\\
\mathfrak{g}_2&=&{\rm tr}[a_2-a_1^2],\\
\mathfrak{g}_3&=&{\rm tr}[a_3-3a_1a_2+2a_1^3],\\
\mathfrak{g}_4&=&{\rm tr}[a_4-4a_1a_3-3a_2^2+12a_1^2a_2-6a_1^4],\\
\mathfrak{g}_5&=&{\rm tr}[a_5-5a_1a_4-10a_2a_3+30a_1a_2^2\nonumber\\&&+20a_1^2a_3-60a_1^3a_2+24a_1^5].
\end{eqnarray}
\end{subequations}
The formula for generic $E_k$ can be written with respect to complete exponential Bell's polynomials as:
\begin{eqnarray}\label{E with respect to g general}
E_k=B_{k}(\mathfrak{g}_1,\mathfrak{g}_2,...,\mathfrak{g}_k).
\end{eqnarray}
%
As before, one can write similar equations for $\tilde{E_k}$ as follows:
\begin{subequations}
\begin{eqnarray}\label{ E tilde and g}
\tilde{E}_1&=&0,\\
\tilde{E}_2&=&\mathfrak{g}_2,\\
\tilde{E}_3&=&\mathfrak{g}_3,\\
\tilde{E}_4&=&\mathfrak{g}_4+3\mathfrak{g}_2^2,\\
\tilde{E}_5&=&\mathfrak{g}_5+10\mathfrak{g}_2\mathfrak{g}_3.
\end{eqnarray}
\end{subequations}
With the following generalization
\begin{eqnarray}\label{ E tilde general XX}
&&\tilde{E}_m+\sum_{j=2}^{m}(-1)^{j-1}(j-1)!B_{mj}(\tilde{E}_1,\tilde{E}_2,...,\tilde{E}_{m-j+1})=\mathfrak{g}_m,\nonumber \\&&\hspace{1cm}m>1.
\end{eqnarray}
%
Consequently, the above formulas show that the quantities $\mathfrak{g}_i$ with $i>2$ represent the fluctuations of the subsystem energy.
 
\subsubsection{Integral representation of the moments}
In this subsection, we introduce an integral representation for the moments that we have calculated in the previous subsection. The new representation
helps to find asymptotic value of $E_2$. Note that similar calculations can be done for the generic XY chain, see Appendix C. However, for the generic
case the integral representation is more complicated.
 A simple calculation shows that
\begin{eqnarray}\label{ E1}
E_1=\frac{2}{\pi}(l-1).
\end{eqnarray}
To calculate $E_2$, we need to evaluate $\mathfrak{g}_2$. However, to calculate $g_2$ we need ${\rm tr} a_2$ and ${\rm tr} a_1^2$. The former one has the following simple form:
\begin{eqnarray}\label{ a2}
{\rm tr} a_2=l-1.
\end{eqnarray}
The later one has more complicated form:
\begin{eqnarray}\label{a12}
{\rm tr} a_1^2=2\sum_{j,k=1}^{l-1}C_{j,k+1}C_{k,j+1}+2\sum_{j=2}^l\sum_{k=1}^{l-1}C_{j,k+1}C_{k,j-1},
\end{eqnarray}
where the correlation matrix $C$ can be written as
\begin{eqnarray}\label{ correlation matrix}
C_{j,k}=\int_{-\pi}^\pi  \frac{dq}{2\pi} n(q)e^{iq(j-k)},
\end{eqnarray}
where $n(q)=1$ for $q<n_c$ and it is zero otherwise. Here, we focus on the half filling $n_c=\frac{\pi}{2}$. Using the above equation and then performing change of variables,
finally, one can write the equation (\ref{a12}) in the following form:
\begin{eqnarray}\label{ a1-2 integral form}
{\rm tr} a_1^2=\frac{2}{\pi^2}\int_0^{\pi}dQ\frac{(\pi-Q+\sin Q)\sin^2[\frac{1}{2}(l-1)Q]}{1-\cos Q}.
\end{eqnarray}
Now, we have an integral formula for $\mathfrak{g}_2$ as follows:
\begin{eqnarray}\label{ g2 integral}
\mathfrak{g}_2=l-1-\frac{l-1}{\pi^2}\int_0^{\pi}dQ(\pi-Q+\sin Q)F_{l-1}[Q],\nonumber\\
\end{eqnarray}
where $F_n[x]=\frac{1}{n}\frac{\sin^2[\frac{nx}{2}]}{\sin^2[\frac{x}{2}]}$ is the Fej\`er kernel. The Fej\`er kernel has the following representation as a sum:
\begin{eqnarray}\label{ Fejer}
F_n[x]=\sum_{k=-n}^n(1-\frac{|k|}{n})e^{ikx}.
\end{eqnarray}
Putting the above equation in the formula (\ref{ g2 integral}) and doing the integral and then performing the sum, we get
\begin{widetext}
\begin{eqnarray}\label{ exact}
\mathfrak{g}_2=\frac{1}{2 \pi ^2}\Big{(}\frac{2 (-1)^l-2 }{l-1}+4 (l-1) \psi ^{(1)}(l-1)-(-1)^l (l-1) \psi ^{(1)}\left(\frac{l}{2}\right)+(-1)^l (l-1) \psi
   ^{(1)}\left(\frac{l+1}{2}\right)\Big{)},
\end{eqnarray}
\end{widetext}
where $\psi ^{(1)}$ is the polygamma function. After expansion around large $l$, we have
\begin{eqnarray}\label{ final}
\lim_{l\to\infty} \mathfrak{g}_2=\frac{2}{\pi^2}+\frac{1}{\pi^2}(\frac{1}{3}+(-1)^l)\frac{1}{l^2}+....
\end{eqnarray}
The above result is consistent with what we expect from conformal field theory calculations.
Similar but much more complicated calculations can be also carried out for $\mathfrak{g}_3$, see Appendix D.
%

\section{ Conclusions}

In this paper, we calculated the full counting statistics of the subsystem energy for an arbitrary free fermion system in its ground state. The provided formula can be also used as the full counting statistics of an
arbitrary quadratic observable. We have calculated exact formulas for the moments as explicit functions of the Hamiltonian parameters. We have also provided a formula for the system in generic state. We then applied our
formulas to the XY chain. In particular, we studied the different behavior of commulants on different parts of the phase diagram. At critical points, we have a power-law decay of the moments but the decay is exponential at
noncritical points. In the case of the XX chain, we provided more refined formulas and showed the validity of the CFT by exact calculations. 

There are many directions that one can expand the current work. For example, it will be very interesting to study the FCS of the energy after quantum quench in one dimension. For some related discussions in this direction see  
Refs.\cite{Silva2008,Arad2016,Markus2015}. It is also important to study the distribution of the subsystem energy in a more direct way (rather than  calculating the generating functions) with
numerical and analytical techniques in quantum spin chains and also bosonic systems. This kind of calculations can also be useful in the field theory context when one is interested in the localization of the energy in a domain.

\textbf{Acknowledgment:}
 We are indebted to Pasquale Calabrese for numerous discussions and suggestions since the beginning of the project.
 We also thank Maurizio Fagotti for many inspiring discussions.
 The work of K.N. is supported by 
 National Science Foundation under Grant No. PHY- 1314295. The work of M.A.R.
is supported in part by CNPq.  M.A.R. thanks ICTP  for hospitality during a period in which part of this work was completed.
\newline
\newline

\section*{Appendix A: First few terms of T matrix expansion} 
\setcounter{equation}{0}
\renewcommand{\theequation}{A\arabic{equation}}

To have a better understanding of the equation ~\ref{sgn}, we show how to obtain the first few terms.
\begin{widetext}
\begin{eqnarray}\label{Tau2even}\
\textbf{T}&=&\mathbb{1}+\bar{\lambda}_{D}\begin{pmatrix}
\textbf{A} & \textbf{B}\\
\textbf{-B} & \textbf{-A}\\
  \end{pmatrix}  +
  \frac{\bar{\lambda}_{D}^2}{2!}\begin{pmatrix}
\textbf{A}^2-\textbf{B}^2 & \textbf{A}\textbf{B}-\textbf{B}\textbf{A}\\
\textbf{A}\textbf{B}-\textbf{B}\textbf{A} & \textbf{A}^2-\textbf{B}^2\\
  \end{pmatrix} + \nonumber \\
  &+&
    \frac{\bar{\lambda}_{D}^3}{3!}\begin{pmatrix}
\textbf{A}^3-\textbf{B}^2\textbf{A}-\textbf{A}\textbf{B}^2+\textbf{B}\textbf{A}\textbf{B} & \textbf{A}^{2}\textbf{B}-\textbf{B}^3-\textbf{A}\textbf{B}\textbf{A}+\textbf{B}\textbf{A}^{2}\\
-\textbf{A}^{2}\textbf{B}+\textbf{B}^3+\textbf{A}\textbf{B}\textbf{A}-\textbf{B}\textbf{A}^{2} & -\textbf{A}^3+\textbf{B}^2\textbf{A}+\textbf{A}\textbf{B}^2-\textbf{B}\textbf{A}\textbf{B}\\
  \end{pmatrix}+...
\end{eqnarray} 
\end{widetext}
The first two terms are easy to study. From equation (\ref{Tau2even}) for $n=0$ we get $\mathbb{1}$. Then for $n=1$ we have two possible terms: 
$\textbf{A}^{1}\textbf{B}^{0}$ and  $\textbf{A}^{0}\textbf{B}^{1}$ as the sum over the power of matrices should satisfy $\sum_{n_i}=n$. 
Moreover, the strings always start with $\textbf{A}$ and ends with $\textbf{B}$. Since $n=1$ is odd, we have to choose $\boldsymbol{\tau}_{1}^{(1)}$ with the condition
$\sum_{j=1}^{k}n_{2j-1}=even$, which implies $n_{1}$ (the power of matrix $\textbf{A}$) to be equal to $0$ and consequently, for $\boldsymbol{\tau}_{1}^{(2)}$ 
we have to choose $n_{1}=1$. So we have $\boldsymbol{\tau}_{1}^{(1)}=\textbf{A}^{0}\textbf{B}^{1}$ and $\boldsymbol{\tau}_{1}^{(2)}=-\textbf{A}^{1}\textbf{B}^{0}$. 
Now, we have to determine the sign which is trivial for this case. 
For $\boldsymbol{\tau}_{1}^{(1)}$ we have $n_1=0$ and $n_2=1$ with $k=2$(total length of AB string) and $s=0$ so that $sgn=(-1)^{[0+0]}=1$ and for $\boldsymbol{\tau}_{1}^{(2)}$, we have
$n_1=1$ and $n_2=0$ which implies $sgn=1$ which is what we expect. Then the other two elements can be obtained from equations (\ref{T11}) and (\ref{T21}).
Similarly, we explain how to extract the terms in elements of $T_{12}$ for $n=3$. For this case, we have $2^{n-1}=4$ different terms. Considering all different $n_{k}$ in equation ~\ref{Tau1odd}, we get,
\begin{eqnarray}\label{T22}\
\textbf{A}^{2}\textbf{B}^1&:&[k=2, n_1=2, n_2=1, s=0], \nonumber \\ &&sgn(k)=(-1)^{[\frac{n_2}{2}+\frac{1-(-1)^{0}}{4}]}=1 \nonumber \\
\textbf{A}^0\textbf{B}^3&:&[k=2, n_1=0, n_2=3, s=0], \nonumber \\ &&sgn(k)=(-1)^{[\frac{n_2}{2}+\frac{1-(-1)^{0}}{4}]}=-1\nonumber \\
\textbf{A}^1\textbf{B}^1\textbf{A}^1\textbf{B}^0&:&[k=4, n_1=1,n_2=1, n_3=1,\nonumber\\&& n_4=0, s=(0,2)],\nonumber \\ &&sgn(k)=(-1)^{([\frac{n_4}{2}+\frac{1-(-1)^{0}}{4}]}\nonumber\\&&^{+[\frac{n_2}{2}+\frac{1-(-1)^{(n_4+n_3)}}{4}])}=-1\nonumber \\
\textbf{A}^0\textbf{B}^1\textbf{A}^{2}\textbf{B}^0&:&[k=4, n_0=1,n_2=1, n_3=2,\nonumber\\&& n_4=0, s=(0,2)] \nonumber \\ &&sgn(k)=(-1)^{([\frac{n_4}{2}+\frac{1-(-1)^{0}}{4}]}\nonumber\\&&^{+[\frac{n_2}{2}+\frac{1-(-1)^{(n_4+n_3)}}{4}])}=1\nonumber \\
\end{eqnarray} 
\newline
\newline

\section*{Appendix B: Bell`s polynomials}
\setcounter{equation}{0}
\renewcommand{\theequation}{B\arabic{equation}}
Here, we summarize some of the properties of the Bell polynomial. 
The partial exponential Bell polynomial is given by the equation
\begin{eqnarray}\label{Bell polynomial}
&&B_{n,k}(x_1,x_2,...,x_{n-k+1})=\nonumber\\ &&\sum\frac{n!}{j_1!j_2!...j_{n-k+1}!}(\frac{x_1}{1!})^{j_1}(\frac{x_2}{2!})^{j_2}...(\frac{x_{n-k+1}}{(n-k+1)!})^{j_{n-k+1}},\nonumber\\
\end{eqnarray} 
where the sum is over all non-negative $j_1,j_2,...,j_{n-k+1}$ in a way that we have $j_1+j_2+...+j_{n-k+1}=k$ and $j_1+2j_2+...+(n-k+1)j_{n-k+1}=n$.
Then the complete  exponential Bell polynomial can be defined as:
\begin{eqnarray}\label{complete Bell polynomial}\
B_n(x_1,x_2,...,x_{n-k+1})=\sum_{k=1}^nB_{n,k}(x_1,x_2,...,x_{n-k+1}).\nonumber\\
\end{eqnarray} 
Now if we define
\begin{eqnarray}\label{complete Bell polynomial 2}\
y_n=\sum_{k=1}^nB_{n,k}(x_1,x_2,...,x_{n-k+1}),
\end{eqnarray}
then we have
\begin{eqnarray}\label{inverse complete Bell polynomial 2}\
x_n=\sum_{k=1}^n(-1)^{k-1}(k-1)!B_{n,k}(y_1,y_2,...,y_{n-k+1}).
\end{eqnarray}
The above formula can be considered inverse relation for Bell polynomials.
\newline
\newline

\section*{Appendix C: Integral representation of $\tilde{E}_2$ for XY chain}
\setcounter{equation}{0}
\renewcommand{\theequation}{C\arabic{equation}}
Here, we write an explicit integral formula of $\tilde{E}_2$  for XY chain. Based on the equation (\ref{Hl2}) we have
\begin{eqnarray}\label{tilde E XY}
\tilde{E}_2=\frac{1}{16}({\rm tr}[\textbf{D}\textbf{D}^{T}]
-{\rm tr}[\textbf{D}^{T}\boldsymbol{G}^{(ba)}\textbf{D}^{T}\boldsymbol{G}^{(ba)}]).
\end{eqnarray} 
To write the integral representation we need to manipulate the second term as follows:
\begin{widetext}

\begin{eqnarray}\label{second term}
\text{tr}[\textbf{D}^{T}\boldsymbol{G}^{(ba)}\textbf{D}^{T}\boldsymbol{G}^{(ba)}]=&&
4h^2\sum_{i,k=1}^l G^{(ba)}_{i,k}G^{(ba)}_{k,i}-4h(1+a)J\sum_{i=1}^{l-1}\sum_{k=1}^l G^{(ba)}_{i,k}G^{(ba)}_{k,i+1}
\nonumber \\&-&4h(1-a)J\sum_{i=1}^{l}\sum_{k=1}^{l-1} G^{(ba)}_{i,k}G^{(ba)}_{k+1,i}+(1+a)^2J^2\sum_{i,k=1}^{l-1} G^{(ba)}_{i,k+1}G^{(ba)}_{k,i+1}\nonumber\\
&+&(1-a)^2J^2\sum_{i,k=1}^{l-1}G^{(ba)}_{i+1,k}G^{(ba)}_{k+1,i}+2(1-a^2)J^2\sum_{i=2}^l\sum_{k=1}^{l-1}G^{(ba)}_{i-1,k}G^{(ba)}_{k+1,i}.
\end{eqnarray} 
\end{widetext}

Note that since the $G^{(ba)}$ matrix is a Toeplitz matrix, one can understand the above formula as the sum of the elements of the pentadiagonal sub-matrix of the matrix $(G^{(ba)})^2$.
Using the above equation and after putting $J=1$, one can simply write the following formula
\begin{widetext}

\begin{eqnarray}
\text{tr}[\textbf{D}^{T}\boldsymbol{G}^{(ba)}\textbf{D}^{T}\boldsymbol{G}^{(ba)}]=\int_0^{2\pi}dx\int_0^{2\pi}dyf(x,y)\frac{(i a \sin (x)-h+\cos (x)) (i a \sin (y)-h+\cos (y))}{\sqrt{a^2 \sin ^2(x)+(\cos (x)-h)^2}
   \sqrt{a^2 \sin ^2(y)+(\cos (y)-h)^2}},
\end{eqnarray}
where 
\begin{eqnarray}
f(x,y)=\frac{2 \left(1-a^2\right) e^{-i l (x+y)} \left(e^{i (l x+y)}-e^{i (l y+x)}\right)^2}{\left(e^{i
   x}-e^{i y}\right)^2}-\frac{4 (1-a) h e^{-i (l-1) (x+y)} \left(e^{i l x}-e^{i l y}\right) \left(e^{i
   (l x+y)}-e^{i (l y+x)}\right)}{\left(e^{i x}-e^{i y}\right)^2}\nonumber\\-\frac{4 (a+1) h e^{-i l (x+y)}
   \left(e^{i l x}-e^{i l y}\right) \left(e^{i (l x+y)}-e^{i (l y+x)}\right)}{\left(e^{i x}-e^{i
   y}\right)^2}+\frac{(1-a)^2 e^{-i (l-1) (x+y)} \left(e^{i (l x+y)}-e^{i (l y+x)}\right)^2}{\left(e^{i
   x}-e^{i y}\right)^2}\nonumber\\+\frac{(a+1)^2 e^{-i (l+1) (x+y)} \left(e^{i (l x+y)}-e^{i (l
   y+x)}\right)^2}{\left(e^{i x}-e^{i y}\right)^2}+\frac{4 h^2 e^{-i (l-1) (x+y)} \left(e^{i l x}-e^{i
   l y}\right)^2}{\left(e^{i x}-e^{i y}\right)^2}.
\end{eqnarray}
\end{widetext}
The above equations for $a=h=0$ produces the formulas for the XX chain introduced in the paper. Although
for generic XY point the above equation is very complicated, there might be especial points that one can handle the above equations analytically. For example, 
for large $l$, the above equation seems to simplify further but we were not able to find the exact expansion for large $l$ in different regimes.
\newline
\newline

\section*{Appendix D: Integral representation of $\tilde{E}_3$ for XX chain}
\setcounter{equation}{0}
\renewcommand{\theequation}{D\arabic{equation}}
In this Appendix, we provide an integral representation for $\mathfrak{g}_3$ defined as: 
\begin{eqnarray}\label{g3}
\mathfrak{g}_3={\rm tr}[a_3-3a_1a_2+2a_1^3].
\end{eqnarray}
The first two terms are easy to calculate. After simple algebra we have

\begin{eqnarray}\label{g3 first two terms}
{\rm tr}[a_3]=\frac{2}{3\pi}l(l+1),\\
{\rm tr}[a_1a_2]=\frac{8}{3\pi}(2l-3).
\end{eqnarray}
The last term $2{\rm tr}[a_1^3]$ is more complicated. It can be written as
\begin{eqnarray}\label{last-term1}
{\rm tr} a_1^3=&&2\sum_{m,n,k=1}^{l-1}C_{n,k+1}C_{k,m+1}C_{m,n+1}\nonumber\\
&+&3\sum_{n=2}^l\sum_{m,k=1}^{l-1}C_{n,k+1}C_{k,m+1}C_{m,n-1}\nonumber\\
&&+3\sum_{k=1}^{l-1}\sum_{n,m=2}^{l}C_{n,k+1}C_{k,m-1}C_{m,n-1}.
\end{eqnarray}
Using the same trick as before and after doing some simplifications, we get
\begin{widetext}
\begin{eqnarray}\label{last-term2}
{\rm tr} a_1^3=\frac{1}{(2\pi)^3}\int_{-\frac{\pi}{2}}^{\frac{\pi}{2}}\int_{-\frac{\pi}{2}}^{\frac{\pi}{2}}\int_{-\frac{\pi}{2}}^{\frac{\pi}{2}}dq_1dq_2dq_3\hspace{4cm}\nonumber\\
(e^{i(q_2-q_1)}-e^{il(q_2-q_1)})(e^{il(q_1-q_3)}-e^{i(q_1-q_3)})(e^{i(q_3-q_2)}-e^{il(q_3-q_2)})\times\nonumber\\
\frac{e^{-iq_1}+e^{-iq_2}+e^{-iq_3}+e^{iq_1}+e^{iq_2}+e^{iq_3}+e^{i(q_1+q_2+q_3)}+e^{-i(q_1+q_2+q_3)}}{(1-e^{i(q_2-q_1)})(e^{i(q_1-q_3)}-1)(1-e^{i(q_3-q_2)})}.
\end{eqnarray}
Then, we can substitiute all of them in equation ~\ref{g3} to get
\begin{eqnarray}\label{last-term2}
\mathfrak{g}_3=\frac{2}{3\pi}l(l+1)-\frac{8}{\pi}(2l-3)+\frac{4}{(2\pi)^3}\int_{0}^{\pi}\int_{0}^{\pi}\int_{0}^{\pi}dq_1dq_2dq_3\hspace{4cm}\nonumber\\
(e^{i(q_2-q_1)}-e^{il(q_2-q_1)})(e^{il(q_1-q_3)}-e^{i(q_1-q_3)})(e^{i(q_3-q_2)}-e^{il(q_3-q_2)})\times\nonumber\\
\frac{\sin q_1+\sin q_2+\sin q_3-\sin[q_1+q_2+q_3]}{(1-e^{i(q_2-q_1)})(e^{i(q_1-q_3)}-1)(1-e^{i(q_3-q_2)})},\hspace{3cm}
\end{eqnarray}
which can be written as
\begin{eqnarray}\label{last-term3}
\mathfrak{g}_3=\frac{2}{3\pi}l(l+1)-\frac{8}{\pi}(2l-3)\hspace{8cm}\nonumber\\
+\frac{4}{(2\pi)^3}\int_{0}^{\pi}\int_{0}^{\pi}\int_{0}^{\pi}dq_1dq_2dq_3\frac{\sin[(l-1)(q_1-q_2)]+\sin[(l-1)(q_3-q_1)]+\sin[(l-1)(q_2-q_3)]}{\sin[q_1-q_2]+\sin[q_3-q_1]+\sin[q_2-q_3]}\times\nonumber\\
(\sin q_1+\sin q_2+\sin q_3-\sin[q_1+q_2+q_3]).\hspace{7cm}
\end{eqnarray}
After some manipulations, one can also get the following form:
\begin{eqnarray}\label{last-term3}
\mathfrak{g}_3=\frac{2}{3\pi}l(l+1)-\frac{8}{\pi}(2l-3)\hspace{8cm}\nonumber\\
+\frac{16}{(2\pi)^3}\int_{0}^{\pi}\int_{0}^{\pi}\int_{0}^{\pi}dq_1dq_2dq_3\frac{\sin[\frac{(l-1)}{2}(q_1-q_2)]\sin[\frac{(l-1)}{2}(q_3-q_1)]\sin[\frac{(l-1)}{2}(q_2-q_3)]}
{\sin[\frac{q_1-q_2}{2}]\sin[\frac{q_3-q_1}{2}]\sin[\frac{q_2-q_3}{2}]}\times\nonumber\\
(\sin[\frac{q_1+q_2}{2}]\sin[\frac{q_3+q_1}{2}]\sin[\frac{q_2+q_3}{2}]).\hspace{7cm}
\end{eqnarray}
\end{widetext}

Although the above equations have nice symmetric forms, it is not easy to find its asymptotic value . Based on numerical calculations it is easy to see that
\begin{eqnarray}\label{g3 final}
\mathfrak{g}_3=c_0+\frac{c_{-3}}{l^3}+...,
\end{eqnarray}
with $c_{-3}$ a bounded but oscillating function. 
\newpage




\end{document}